\documentclass[twocolumn,aps,prl,superscriptaddress]{revtex4}
\usepackage[utf8]{inputenc}
\usepackage{epsf,graphicx}
\usepackage{multirow}
\usepackage{amsmath,gensymb,amssymb}
\usepackage{lipsum}
\usepackage{dcolumn}
\usepackage{bm}
\usepackage{hyperref}
\usepackage{natbib}
\usepackage{color}

\begin{document}

\title{Development of strongly nonlinear structures at the charged boundary of a non-conducting liquid in an electric field}

\author{N.M. Zubarev}
\email{nick@iep.uran.ru}
\affiliation{Institute of Electrophysics, Russian Academy of Sciences, Yekaterinburg, 620016 Russia}
\affiliation{Lebedev Physical Institute, RAS, 119991 Moscow, Russia}
\date{\today}

\author{E.A. Kochurin}
\email{kochurin@iep.uran.ru}
\affiliation{Institute of Electrophysics, Ural Division, Russian Academy of Sciences, Yekaterinburg, 620016 Russia}
\affiliation{Skolkovo Institute of Science and Technology, 121205, Moscow, Russia}

\begin{abstract}
Direct numerical simulation of the strongly nonlinear stages of instability development for a non-conducting liquid with a charged free surface in a normal electric field is performed. It is demonstrated that two main stages of the instability can be distinguished: an initial stage, during which dimples appear on the surface, and a developed stage, during which these dimples transform into expanding bubbles. The bubble size increases with increasing applied field, despite the fact that the scale corresponding to the dominant mode of the instability decreases.
\end{abstract}

\maketitle

An electric field directed perpendicular to the boundary of a liquid with a surface charge has a destabilizing effect on it. The Tonks-Frenkel instability \cite{z1,z2}, which arises at the surface of a perfectly conducting liquid (e.g., molten metal) in a sufficiently strong external electric field, has been studied most thoroughly. In this case, the field does not penetrate the liquid due to screening by the free electric charge induced on its surface. The opposite situation is also possible, which is the subject of this paper, when an electric field is present in the liquid but absent outside it. This could be a non-conducting liquid with a free surface electric charge, such as liquid helium with electrons localized above its surface \cite{z3,zz3}, or the interface between a non-conducting liquid and a significantly lighter conducting medium. An example of the latter situation is the boundary between a dielectric liquid and a plasma during the ignition of partial electric discharges in gas bubbles rising in the liquid \cite{z4}.

The linear stages of instability development of a flat liquid boundary in both situations (when the field is present only outside the liquid or only inside the liquid) are identical: they are described by the same dispersion law
\begin{equation} \label{eq1}
\omega ^{2} =\left(\alpha /\rho \right)k^{3} -\left(\varepsilon _{0} \varepsilon E^{2} /\rho \right)k^{2},
\end{equation}
where $\omega$ is the frequency, $k$  is the wavenumber, $\alpha$ is the surface tension coefficient, $\rho$ is the liquid density, $\varepsilon_{0}$ is the electric constant, $E$ and $\varepsilon$ are the field strength and permittivity outside the liquid (in the first case) or in the liquid (in the second case).

It is interesting to compare the nonlinear stages of instability development, i.e., the stages at which the surface inclination angles cease to be small. At these stages, the difference between the situations under discussion will become evident, namely that the electric field is present on different sides of the boundary. Since the instability of the liquid boundary results from the interaction of the applied field and the surface charge (the charge is drawn in by the field), surface disturbances directed outward in the first case and inward in the second will grow more rapidly. This determines the tendency for protrusions (bumps) to form on the surface of a conducting liquid and depressions (dimples) for a non-conducting liquid.

What will be the behavior of the fluid at strongly nonlinear stages of instability? Some clues can be given by the general properties of the original equations of motion, which are also reflected in the dispersion law \eqref{eq1}. The equations are invariant with respect to the dilations $E\to cE$, $\lambda \to c^{-2} \lambda $, $\tau \to c^{-3} \tau $ (here $c$ is an arbitrary dimensionless constant), i.e., the characteristic spatial ($\lambda $) and temporal ($\tau $) scales of the problem change with the field following the power laws $E^{-2} $ and $E^{-3} $, respectively. Indeed, a key role in the development of instability is played by the so-called dominant mode corresponding to the maximum of its growth rate, i.e., the extremum $\mbox{Im}(d\omega /dk)=0$ in the dispersion law. As can be easily found from \eqref{eq1}, we have $k_{d} =2\varepsilon \varepsilon _{0} E^{2} /3\alpha $ and $\gamma _{d} =\mbox{Im}\,\omega (k_{d} )\propto E^{3} $ for the wavenumber and the growth rate of the dominant instability mode.
Thus, the characteristic scales of the problem are related to the field $E$ as $\lambda \propto k_{d}^{-1} \propto E^{-2}$ and
$\tau \propto \gamma _{d}^{-1} \propto E^{-3}$.
Moreover, it was shown in \cite{z5,z6,zz6,zzz6} for the Tonks-Frenkel instability that the process of its development is scale-invariant: self-similar solutions corresponding to the scaling $\lambda \propto \tau ^{2/3} $ are realized. They are responsible for the formation of conical cusps on the liquid surface. The dynamics of the liquid near the forming cones, i.e., at strongly nonlinear stages of instability, will be determined not by the external but by the local field $E_{loc} $. The relations for the scales $\lambda \propto E_{loc}^{-2} $ and $\tau \propto E_{loc}^{-3} $ are then valid linking the unlimited sharpening of the surface in a finite time (the quantities $\lambda $ and $\tau $ become zero) and the growth of the local field.

As will be shown below, these considerations, which allow us to adequately describe the behavior of a conducting liquid in an electric field, are not applicable to the case of a non-conducting liquid, which is of interest to us. Contrary to the estimate for the dominant mode scale
$\lambda _{d} = 2\pi /k_{d} \propto E^{-2}$,
the characteristic size of the forming structures increases with increasing $E$, i.e., they are determined by other factors. This qualitatively distinguishes the compared situations despite the identical descriptions in the linear approximation and the same scaling.

Let us consider a two-dimensional plane potential flow of an ideal incompressible fluid with a free boundary $y=\eta (x,t)$ in a uniform electric field of strength $E$. In the unperturbed state $\eta =0$, i.e., the fluid occupies the half-plane $y\le 0$. Here, a Cartesian coordinate system $\{ x,y\} $ is used, for which the $y$-axis is codirectional with the applied electric field, and the $x$-axis coincides with the unperturbed boundary. The fluid is assumed to be non-conducting (dielectric) with permittivity $\varepsilon $. A free surface charge is distributed along its boundary ensuring its equipotentiality. We assume that the surface charge density in the unperturbed state is $\sigma =-\varepsilon _{0} \varepsilon E$, which guarantees complete screening of the electric field above the fluid. The potentials of fluid velocity $\phi $ and electric field $\varphi$ in the region $y<\eta (x,t)$ satisfy the Laplace equations $\nabla ^{2} \phi =0$ and $\nabla ^{2} \varphi =0$. On the liquid surface $y=\eta (x,t)$, the dynamic and kinematic boundary conditions are satisfied:
\[
\phi _{t} +\frac{1}{2} (\nabla \phi )^{2} =\frac{\alpha }{\rho } K-\frac{\varepsilon _{0} \varepsilon }{2\rho } (\nabla \varphi )^{2} ,
\]
\[
\eta _{t} +\eta _{x} \phi _{x} =\phi _{y} ,
\]
where $K=\eta _{xx} (1+\eta _{x}^{2} )^{-3/2}$ is the local curvature of the boundary. The equations of motion are closed by the conditions $\nabla \phi \to 0$ and $\nabla \varphi \to \{ 0,-E\} $ at the depth $y\to -\infty $, and by the condition $\varphi =0$ at the boundary $y=\eta (x,t)$. In the case of harmonic perturbations of the surface $\eta \propto \exp (ikx-i\omega t)$ of small amplitude (i.e. $\left|\eta _{x} \right|\ll 1$), these equations lead to the dispersion law \eqref{eq1}.

We will assume that the functions $\phi $, $\varphi $, and $\eta $ are periodic along the $x$-axis with some given period $\lambda _{0} $. We will switch to dimensionless notation, using the corresponding wavenumber $k_{0} =2\pi /\lambda _{0} $ and frequency $\omega _{0} =(\alpha k_{0}^{3} /\rho )^{1/2} $:
\[
\tilde{y}=yk_{0} ,\quad \tilde{x}=xk_{0} ,\quad \tilde{\eta }=\eta k_{0},
\]
\[
\tilde{t}=t\omega _{0} ,\quad \tilde{\phi }=\phi k_{0}^{2} /\omega _{0} ,\quad \tilde{\varphi }=\varphi k_{0} /E.
\]
In dimensionless form, the dispersion relation \eqref{eq1} takes the form (here and below, the ``$\sim$'' signs are omitted for convenience):
\[
\omega ^{2} =k^{3} -\beta ^{2} k^{2} ,
\]
where the dimensionless parameter $\beta ^{2} =\varepsilon _{0} \varepsilon E^{2} /\alpha k_{0} $ is introduced. Since $\beta \propto E$, it can be considered to characterize the strength of the applied electric field. The parameter $\beta $ can also be interpreted in terms of the characteristic scales of the problem. Two scales were mentioned above. The scale $\lambda _{d} $ corresponds to the dominant instability mode (perturbations of this length grow most rapidly). The scale $\lambda _{0} $ has the meaning of the spatial period of the problem, i.e., it actually determines the length of the initial boundary perturbation. The relation $\beta ^{2} =3k_{d} /(2k_{0} )\propto \lambda _{0} /\lambda _{d} $ is valid according to which the control parameter of the problem $\beta $ is determined by the ratio of these scales.

In the absence of an electric field ($\beta =0$), the above system of equations describes the propagation of capillary waves. Under the influence of the field, as can be seen from the linear dispersion law for $k=1$, the value of $\omega ^{2} $ becomes negative and, consequently, the liquid boundary loses stability for $\beta >1$. The near-critical behavior of the system ($\beta \approx 1$) can easily be described analytically in a weakly nonlinear approximation when the amplitude of the boundary perturbation is considered small but finite. Following by analogy with work \cite{z7} which considered the behavior of a conducting liquid in an electric field we find for the shape of the boundary:
\[
\eta (x,t)=-A(t)\cos x-(1/4)A^{2} (t)\cos 2x+\ldots ,
\]
where the amplitude $A$ satisfies an ordinary differential equation (ODE)
\begin{equation} \label{eq2}
A_{tt} =(\beta ^{2} -1)A+A^{3} /8
\end{equation}
with nontrivial stationary solutions $A^{2} =8(1-\beta ^{2} )$ corresponding to a subcritical ``pitchfork'' bifurcation. According to \eqref{eq2}, nonlinearity plays a destabilizing role causing a rigid loss of stability of the system. For $\beta <1$, when the flat boundary of the fluid is stable in the linear approximation, instability develops if at the initial instant of time the following holds:
\begin{equation} \label{eq3}
A_{t}^{2} +(1-\beta ^{2} )A^{2} >4(\beta ^{2} -1)^{2} +A^{4} /16,
\end{equation}
i.e., when the values $A(t)$ and $A_{t}(t)$ are sufficiently large.

As can be seen from \eqref{eq2}, a distinctive feature of the instability under consideration is its explosive nature: the amplitude grows infinitely with an asymptotic behavior of $A\propto 1/(t_{c} -t)$, where $t_{c} $ is the moment of singularity formation. It is clear that the weakly nonlinear model \eqref{eq2} here goes beyond its applicability, and to describe the developed stages of instability it is necessary to solve the original equations of motion directly. An important feature of these equations, which was first noted and used in \cite{z8,z9}, is that the domains of definition of the harmonic potentials $\phi $ and $\varphi $ coincide and, as a consequence, the method of dynamic conformal transformations developed for describing capillary-gravity waves \cite{z10,z11,z12} can be applicable.

Let us perform a conformal transformation of the region occupied by the liquid into the half-plane $v\le 0$ in parametric variables $\{ u,v\} $. The boundary of the liquid will correspond to the line $v=0$, and the function $\eta (x,t)$ is given in parametric form by the relations
\[
\eta =Y(u,t),\quad x=X(u,t)=u-\hat{H}Y(u,t),
\]
where $\hat{H}$ is the Hilbert transform, defined in Fourier space as $\hat{H}f_{k} =i\,\mbox{sign}(k)f_{k}$. We also introduce the function $\Psi (u,t)$, which defines the velocity potential at the fluid boundary. In conformal variables, the original two-dimensional problem reduces to a pair of one-dimensional equations for the motion of the fluid boundary,
\[
\Psi _{t} =\frac{1}{2J} \left[(\hat{H}\Psi _{u} )^{2} +\Psi _{u}^{2} \right]+\Psi _{u} \hat{H}\left[\frac{\hat{H}\Psi _{u} }{J} \right]-
\]
\begin{equation} \label{eq4}
-\frac{\beta ^{2} }{2J} -\frac{X_{u} Y_{uu} -Y_{u} X_{uu} }{J^{3/2}},
\end{equation}
\begin{equation} \label{eq5}
Y_{t} =Y_{u} \hat{H}\left[\frac{\hat{H}\Psi _{u} }{J} \right]-X_{u} \frac{\hat{H}\Psi _{u} }{J},
\end{equation}
where $J=X_{u}^{2} +Y_{u}^{2} $ is the Jacobian of the transformation. The term $\beta ^{2} /2J$ is responsible for the electrostatic pressure. Note that the equations taking into account the influence of the electric field were first derived in work \cite{z13} (see also \cite{z14}) in a form unresolved with respect to time derivatives. For numerical modeling of the instability development, it is more convenient to use the representation \eqref{eq4}, \eqref{eq5}. Moreover, to improve the accuracy of the numerical solution, it is advisable to pass to the Dyachenko variables \cite{z11,z12} $V=i\Phi _{u} /Z_{u} $ and $R=1/Z_{u} $, where the complex functions $Z=X+iY$ and $\Phi =\Psi +i\hat{H}\Psi $ are used. In terms of the variables $V$ and $R$, the equations \eqref{eq4} and \eqref{eq5} take the form:
\begin{equation} \label{eq6}
V_{t} =i(UV_{u} -D_{u} R)-2Q^{2} \hat{P}(Q_{u} \bar{Q}-\bar{Q}Q_{u} )_{u},
\end{equation}
\begin{equation} \label{eq7}
R_{t} =i(UR_{u} -U_{u} R),
\end{equation}
where $U=\hat{P}(V\bar{R}+\bar{V}R)$, $D=\hat{P}(V\bar{V}+\beta ^{2} R\bar{R})$, $Q=R^{1/2} $ are denoted, the projection operator $\hat{P}=(1+i\hat{H})/2$ is introduced, and the bar above the symbols denotes complex conjugation.

An interesting feature of the problem under consideration is that if surface tension is neglected, that is, if the evolution of the fluid is considered under the influence of electrostatic forces alone, the equations of motion can be substantially simplified. Indeed, if we discard the last term on the right-hand side \eqref{eq6} responsible for capillary effects, the resulting system
\[
V_{t} =i(UV_{u} -D_{u} R),\quad \quad R_{t} =i(UR_{u} -U_{u} R)
\]
after substituting $V=\beta (R-1)$ is reduced to a single equation
\[
R_{t} =2i\beta \hat{P}\left(R_{u} \hat{P}(R\bar{R})-R\hat{P}(R\hat{R})_{u} \right).
\]
In turn, it is reduced to the well-known Laplacian growth equation (LGE) \cite{z15,z16} $\mbox{Im}(\bar{G}_{t} G_{u} )=\beta $, where $G(u,t)=Z(u,t)-i\beta t$. The possibility of reducing the problem to the LGE was discovered in the work \cite{z8}, and the integrability of the LGE \cite{z17,z18,z19,z20,z21} made it possible to construct exact nontrivial solutions for the boundary evolution at strongly nonlinear stages of instability \cite{z8,z9,z22}.

The simplest periodic solution of the LGE has the form
\begin{equation} \label{eq8}
Y(u,t)=-A(t)\cos u-A^{2} (t)/2,
\end{equation}
where the amplitude $A$ satisfies the ODE
\begin{equation} \label{eq9}
A_{t} =\beta A/(1-k^{2} A^{2} ).
\end{equation}
The evolution of the free surface corresponding to the solution \eqref{eq8} and \eqref{eq9} is shown in Fig.~1 (for definiteness, $\beta = 1$ is chosen). It is evident that a singularity (a pointed dimple) forms on the boundary over a finite time, in the neighborhood of which $\eta (x)-\eta(0)\propto \left|x\right|^{2/3} $. At a singular point, the electrostatic pressure $\beta /2J$ and the boundary curvature $K$ become infinite: see the inset in Fig.~1.

\begin{figure}[t]
	\centering
	\includegraphics[width=0.9\linewidth]{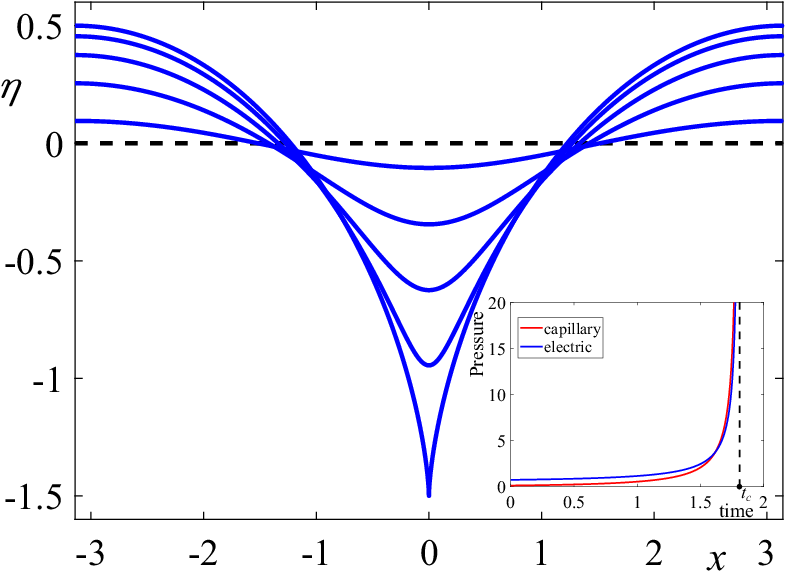}
	\caption{ Fig. 1. (Color online) Evolution of the free surface within the exact solution \eqref{eq8} and \eqref{eq9} for $\beta =1$ with the initial condition $A(0)=0.1$. The moment of collapse $t\approx 1.8076$. The inset shows the time dependencies at the point $x=0$ of the electrostatic pressure (blue line) and the capillary pressure (red line), which was not taken into account in the calculations.}
	\label{fig_1}
\end{figure}

It is clear that the singularity demonstrated in Fig.~1 can only arise if capillary effects are neglected. For the complete system \eqref{eq6} and \eqref{eq7}, the electrostatic and capillary pressures at $\beta = 1$ will be comparable. The question arises: what behavior will the system exhibit at developed stages of instability when surface tension is taken into account? To answer this question, we will make direct numerical simulations of the liquid boundary evolution based on the model system \eqref{eq6} and \eqref{eq7} using pseudo-spectral methods. This numerical approach combined with the conformal mapping method has shown high efficiency in describing electrohydrodynamic fluid flows \cite{zuko, flam1, gao, dop}, flows with constant vorticity \cite{flam2, flam3, flam4}, water waves \cite{ruban1, ruban2, korot,korot1}, and wave turbulence \cite{koch1, koch2}.

We choose initial conditions in the form of periodic perturbations of the surface and velocity field coinciding with those used in the exact solutions \eqref{eq8} and \eqref{eq9}:
\begin{equation} \label{eq10}
Y(u,0)=-A\cos u,\quad \Psi (u,0)=\beta Y(u,0).
\end{equation}
The surface evolution obtained by numerical integration of \eqref{eq6} and \eqref{eq7} at $\beta =1$, i.e., at the threshold electric field value for instability development, is shown in Fig.~2. It is evident that the behavior of the liquid differs radically from that described by the exact solution \eqref{eq8} and \eqref{eq9} presented in Fig.~1. Instead of an infinite sharpening of the dimple, it is transformed into a bubble. The calculation is interrupted at the moment $t\approx 13.3$, when the topology of the problem changes qualitatively: self-intersection of the liquid boundary is observed corresponding to the detachment of a charged bubble.

\begin{figure}[t]
	\centering
	\includegraphics[width=0.9\linewidth]{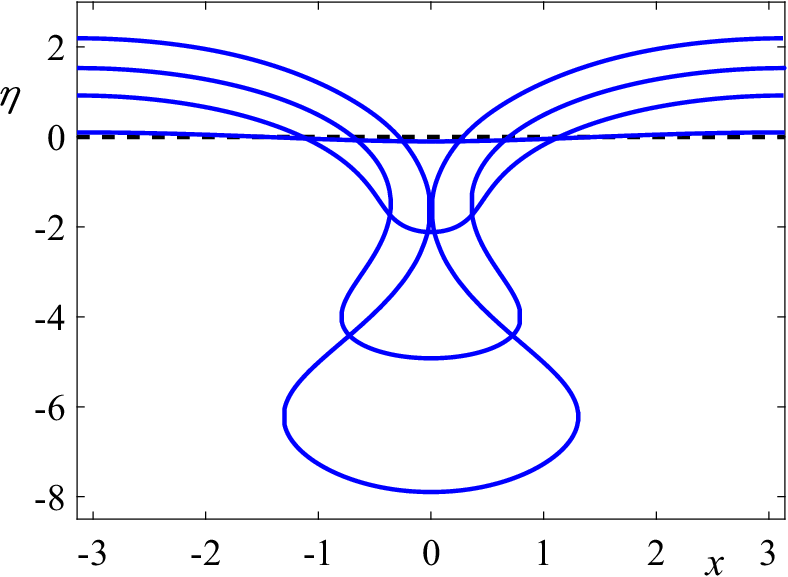}
	\caption{ Fig. 2. (Color online) Evolution of the liquid surface obtained as a result of numerical solution of the model \eqref{eq6} and \eqref{eq7} for $\beta =1$ and the initial condition \eqref{eq10} with $A=0.1$. Successive moments of time $t=0, \, 7.7,\, 10.0,\, 13.3$ are shown.}
	\label{fig_2}
\end{figure}

\begin{figure}[t]
	\centering
	\includegraphics[width=0.9\linewidth]{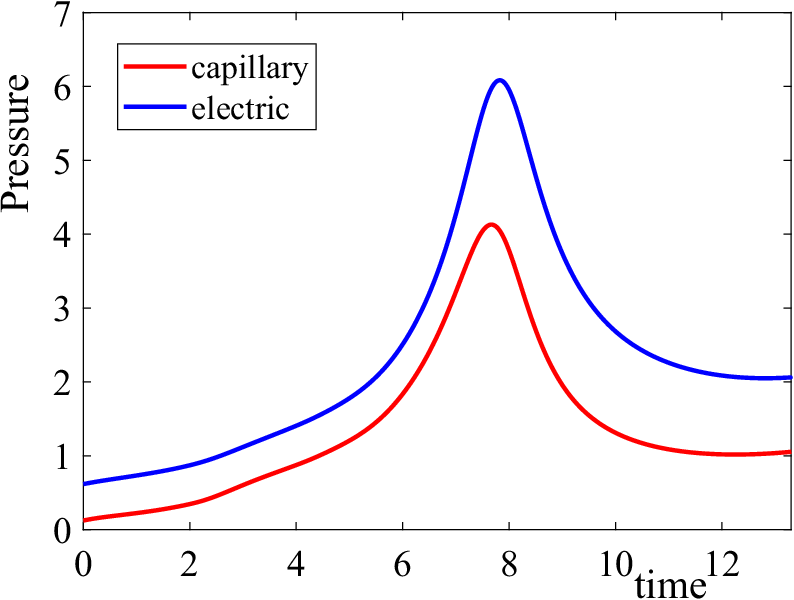}
	\caption{ Fig. 3. (Color online) Time dependencies of electrostatic pressure (blue line) and capillary pressure (red line) at point $x=0$ for $\beta =1$ and initial condition \eqref{eq10} with $A=0.1$.}
	\label{fig_3}
\end{figure}

The instability development process can be divided into two stages as illustrated by the time dependencies of the electrostatic and capillary pressures at the point $x=0$ shown in Fig.~3. Both quantities initially increase, which qualitatively coincides with their behavior for the exact solution \eqref{eq8} and \eqref{eq9}, shown in the inset of Fig.~1. At this stage of instability development, a dimple forms at the liquid boundary. However, unlike the situation shown in Fig.~1, unlimited sharpening of the dimple does not occur. The electrostatic and capillary pressures reach a maximum at $t\approx 7.7$ and then begin to decrease. This reflects a qualitative change in the behavior of the system: the next stage of instability development begins, during which a bubble forms and expands. It ends with the detachment of the bubble at $t\approx 13.3$.

Thus, the instability stage in which a dimple forms in the surface (the surface curvature increases) is followed by a stage in which it transforms into an expanding bubble (the curvature decreases). Let us consider what determines the spatial scale of the forming structure. A natural choice for this scale is the bubble size at the moment of its detachment. At first glance, it should correlate with the wavelength of the dominant instability mode and due to the relation $\lambda _{d} \propto E^{-2} $ decrease with increasing field strength tending to zero as $E\to \infty $. Singular solutions of the \eqref{eq8} and \eqref{eq9} types corresponding to the strong field limit (capillary effects are considered negligible compared to electrostatic ones) fit well into this logic. For these solutions, the characteristic spatial scale -- the radius of curvature of the surface -- becomes zero at the final stage of instability.

However, this assumption is refuted by the results of numerical calculations of the fluid boundary evolution for various values of the control parameter $\beta $ (or the field $E$ in the original notation). The calculations were performed for initial conditions \eqref{eq10} in the range $0.75\le \beta ^{2} \le 1.2$, i.e., for both subcritical and supercritical values of $\beta $. In the supercritical case ($\beta >1$), the initial amplitude of the surface perturbation was taken to be $A=0.1$. In the subcritical case ($\beta <1$), when the flat surface is stable with respect to small perturbations and its destabilization requires the fulfillment of the condition \eqref{eq3}, the initial amplitude increased to $A=0.8, \, 0.5, \, 0.3$ for $\beta ^{2} =0.75, \, 0.8, \, 0.9$, respectively. The calculations were interrupted when the liquid boundary self-intersected: see the inset in Fig.~4. At this moment, the average radius of the bubble being formed was calculated based on its length; the formula was
\[
\langle R\rangle =(2\pi )^{-1}\!\!\int\!\!\sqrt{X_{u}^{2}+Y_{u}^{2}}\, du.
\]

\begin{figure}[t]
	\centering
	\includegraphics[width=0.9\linewidth]{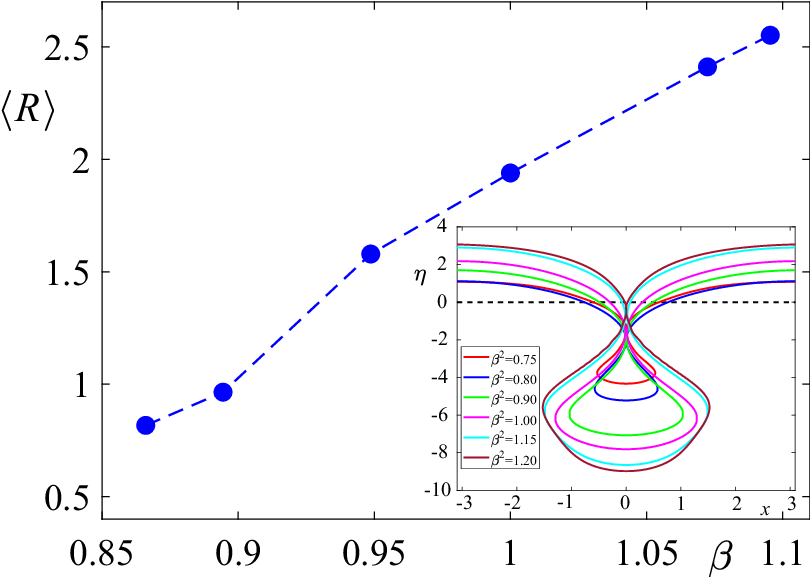}
	\caption{ Fig. 4. (Color online) Average radius of the formed bubble depending on the parameter $\beta $. The inset shows the shapes of the liquid surface at the moment of bubble detachment for $\beta ^{2} = 0.75, \, 0.8, \, 0.9, \, 1, \, 1.15, \,1.2$.}
	\label{fig_4}
\end{figure}

The calculation results are shown in Fig.~4. It can be seen that the bubble radius increases monotonically with increasing $\beta $ (or $E$), i.e., it is not determined by the scale $\lambda _{d} $, which decreases with the field as $E^{-2} $. What is the reason for this unexpected behavior of the system? We attribute the bubble expansion to the influence of repulsive forces between free electric charges that flow into the dimple. The surface area from which the charge accumulates is determined not by the scale $\lambda _{d} $, but by the scale $\lambda _{0} $, i.e., by the period of the initial perturbation of the boundary. We will show that it is precisely this circumstance that determines the nature of the dependence of the bubble radius on the applied field.

Let us estimate the bubble radius based on several assumptions. First, for simplicity, we will assume that the bubble is round with radius $R$. Second, we will assume that all the electric charge belonging to the region $-\lambda _{0} /2\le x\le \lambda _{0} /2$ has flowed into it. If we assume that it is uniformly distributed over the bubble surface, then the field strength at its boundary $E_{b} $ is estimated from the charge conservation law as $2\pi RE_{b} =\lambda _{0} E$. Finally, third, we will assume that the bubble radius corresponds to the balance of electrostatic and capillary pressures at its boundary: $\alpha /R=\varepsilon \varepsilon _{0} E_{b}^{2} /2$. From here we find:
\begin{equation} \label{eq11}
R=\varepsilon \varepsilon _{0} E^{2} \lambda _{0}^{2} /(8\pi ^{2} \alpha ),
\end{equation}
or, in dimensionless form, $R=\beta ^{2} /2$. Thus, the bubble radius increases monotonically with increasing parameter $\beta $ (i.e., in fact, the field $E$), which is in qualitative agreement with the dependence shown in Fig.~4. Note that formula \eqref{eq11} can be rewritten as $R=3\lambda _{0}^{2} /(8\pi \lambda _{d} )$, i.e., the size of the emerging structures is determined by both spatial scales of the problem -- the period of the initial disturbance $\lambda _{0} $ and the wavelength of the dominant mode $\lambda _{d} $.

Thus, in this paper, we have described the strongly nonlinear dynamics of the boundary of a non-conducting liquid with a free surface charge within a physical model that takes into account both electrostatic and capillary pressures. It has been established that two main stages can be distinguished in the development of boundary instability: an initial (weakly nonlinear) stage, during which a dimple appears in the surface, and a developed (strongly nonlinear) stage, during which the dimple transforms into an expanding bubble. The presence of the bubble expansion stage fundamentally distinguishes the problem under consideration from the related problem of the behavior of a perfectly conducting liquid in an electric field. The analogue of the bubble formation process for this problem would be the droplet formation process \cite{z7}; however, droplet expansion like bubble expansion is impossible due to the incompressibility of the liquid. Consequently, despite the fact that the linear dispersion laws for both problems coincide, and the equations of motion are invariant under identical scaling transformations, the final stages of instability are qualitatively different. If for a conducting liquid the characteristic scale of the forming structures decreases with the growth of the applied field in accordance with the similarity laws of the original equations, then for a non-conducting liquid the scale -- the size of the forming bubbles -- increases.

{\bf FUNDING}

The work was carried out within the framework of the grant of the Russian Science Foundation No. 23-71-10012.

{\bf CONFLICT OF INTEREST }

The authors of this work declare that they have no conflict
of interest

\end{document}